\documentclass[twocolumn]{elsart}
\usepackage{graphicx}

\usepackage{amssymb}
\def    \be             {\begin{equation}}
\def    \ee             {\end{equation}}
\def    \ba             {\begin{eqnarray}}
\def    \ea             {\end{eqnarray}}

\def    \=              {\;=\;}
\def    \frac           #1#2{{#1 \over #2}}
\def    \mh             {\mbox{$m_H$}}
\def \lsim{\mathrel{\vcenter
     {\hbox{$<$}\nointerlineskip\hbox{$\sim$}}}}
\def \gtrsim{\mathrel{\vcenter
     {\hbox{$>$}\nointerlineskip\hbox{$\sim$}}}}
\def\gev{\mbox{GeV}}
\def\mh{\mbox{$m_H$}}
\def\btau{B_\tau}

\begin{document}

\begin{frontmatter}



\title{The future of particle physics \thanksref{conf}}
\thanks[conf]{Opening talk at the 10$^{th}$ Pisa meeting on advanced
  detectors, La Biodola, May 22-27 2006.}


\author{Michelangelo L. Mangano}

\address{TH Unit, PH Department, CERN, 1211 Geneva, Switzerland}

\begin{abstract}
I review the prospects for future progress in accelerator-based 
particle physics
\end{abstract}

\end{frontmatter}

\section{Introduction}
\label{sec:intro}
To discuss the future it is useful to review the past first. By 1973
the theoretical foundations of the Standard Model (SM) were fully
established, the last ones being the proof of renormalizability and
unitarity of the $SU(2)\times U(1)$ Yang-Mills Lagrangian with the
Higgs mechanism of EW symmetry breaking (EWSB), the discovery of
asymptotic freedom and the ensuing proposal of QCD as the gauge theory
of strong interactions, and, finally, the Kobayashi-Maskawa (KM)
description of CP violation with a fermionic 3-family structure. After
1973 followed over 30 years of consolidation, whose main ingredients
are summarized as follows: (i) theoretical technical advances
(development of techniques for more and more accurate calculations,
and lattice gauge theories to deal with the non-perturbative aspects
of strong interactions); (ii) experimental verification of the SM
spectrum, with the discovery of the new fermions (charm, plus all
members of the third generation) and of the predicted gauge bosons
(the gluon, $W$ and $Z$); and (iii) experimental verification of the
SM dynamics, with the measurement and test of EW radiative
corrections, of the running of $\alpha_s(Q)$, and, finally, the
confirmation of the KM model of quark mixings and CP violation (the
measurement of direct CP violation in $K$ decays, and the recent
successful tests for the third generation performed at LEP/SLC, the
Tevatron and, most compelling, at the $B$-factories).

Those who claim that nothing interesting has happened in particle
physics in the past 30 years should think twice. The formulation and
consolidation of the SM is a monumental scientific achievement, with
parallels only in the discovery of Maxwell's theory of electromagnetism,
of special and general relativity, and of quantum mechanics. These
past 30 years will be recorded in history as a milestone in the
development of our understanding of Nature.

After 1973, theoretical progress has been mostly driven by theory
itself, rather than by data. The push came from the need of a better
understanding of several issues left open by the SM: (i) identifying the deep
origin of EWSB, (ii) of the gauge structure, (iii) of the family
structure and, last but not least, (iv) the understanding of quantum
gravity and of the vanishing of the cosmological
constant (which, most recently, turned out to be true only
approximately, making the puzzle even more challenging). 
The milestones that emerged from these speculations
include: 
\begin{itemize}
\item GUTs (grand unified theories, 1974), to extend
  Maxwell's unification to all gauge interactions;
\item Supersymmetry (SUSY, 1974), to complete the set of mathematically
  allowed symmetries of space-time;
\item The see-saw mechanism (1977), to provide a dynamical explanation of the
  smallness of neutrino masses;
\item Technicolour (1979), to provide a dynamical framework for EWSB and the
  Higgs mechanism;
\item Inflation (1980), to explain the flatness of the Universe;
\item Superstrings (1984), to provide a consistent theory of quantum
  gravity and a possible Theory of Everything;
\item Large-scale extra dimensions, to provide a natural explanation
  of the large difference between the strength of gauge and
  gravitational interactions.
\end{itemize}
In addition to the above, theoretical physics has witnessed the
establishment and consolidation of a Standard Model of cosmology,
based on general relativity, particle physics, and inflation, capable
of explaining, among others, properties of the Universe as diverse as
the nuclear abundances, the fine structure of the microwave
background radiation, and the formation of large-scale structures.

Since 1973, experimental particle physics has been mostly occupied
with verifying the SM, as mentioned above, and attempting to find
traces of the new theoretical ideas that were being put forward:
proton decay or neutron oscillations in the case of GUTs; signatures
of sparticles in the case of SUSY; neutrino masses or mixings;
signatures of extra-dimensions (deviations from Newton's law or
graviton emission in hard collisions); and more. With the exception of the
discovery of neutrino oscillations, nothing compelling has
unfortunately emerged as yet. This resembles the most frustrating
among the scenarios envisaged by Glashow in a seminar with a title
similar to mine given almost 30 years ago~\cite{Glashow:1979nm}. 
Are we therefore destined
to be stuck forever with the SM? Why should we expect that something
new and exciting will happen soon, with the accelerator and
experiments that are operating or about to start?

The are two sets of reasons that justify the expectation that
something new and exciting is just about to happen: the first set
relies on theoretical prejudice, the second on experimental
facts. We shall now examine these two complementary viewpoints.

\section{The theoretical wisdom}
\subsection{Electroweak symmetry breaking}
The required step towards future progress in particle physics is today
the observation of the Higgs boson, which should lead to the beginning
of a clarification of the EWSB mechanism. From the theoretical point of
view, Higgs boson searches at the LHC will provide non trivial information
regardless of the outcome. If the SM description of EWSB is correct
(and if the LHC and the experiments perform as expected, something
we'll give for granted), the
observation of the Higgs is guaranteed. The SM fits of the current EW
data firmly predict at 95\%CL that $m_H\lsim 200$ GeV; 
the direct LEP limit, valid within the
SM, says that $m_H\gtrsim 114$ GeV. In the mass range $114<m_H(\gev)<200$,
ATLAS and CMS promise a discovery with an integrated luminosity
between 1 and 10 fb$^{-1}$. If this does not happen, the SM is in
trouble. Whether the Higgs is not seen because it decays to final
states with small detection efficiency, or because the production
rates are much smaller than predicted, in all cases this would point
to physics BSM, since production rates and decay modes and BRs are
uniquely predicted with good accuracy by the SM. A SM-like Higgs with
a mass of several hundred GeV, visible at the LHC for masses up to
about one TeV, would also create problems to the SM, since such a
large mass would
conflict with the EW measurements. Complete lack of a Higgs resonance
below the TeV, finally, would also be a clear indication of new
physics, because of a violation of perturbative unitarity in WW
scattering at high energy. In the context of standard 4-dimensional
field theories this could only be circumvented
by the appearance of resonances in gauge boson scattering around the
TeV, yet another interesting new phenomenon.

Contrary to previous accelerators, like LEP2 or the Tevatron, the LHC
will therefore be able to conclusively answer the question of whether
or not
nature is consistent with the SM description of
EWSB. Regardless of the outcome, one of the most long-awaited
questions in particle physics will soon be answered.
Even if the Higgs will appear to behave like in the SM (i.e. its mass
will be consistent with the current bounds and its production and
decay properties will match those predicted by the SM), there is no
guarantee that no other underlying phenomena are at work, and
therefore in all cases a more complete exploration of the EWSB
dynamics will need to be carried out. In particular, EWSB as described in
the SM opens a major theoretical puzzle, discussed in the next section,
which strongly calls for physics BSM.

\subsection{The hierarchy problem}
Radiative corrections
induced by the coupling with the top quark generate a shift of the
Higgs mass squared:
\be \label{eq:deltamh}
\Delta m_H^2 = \frac{6G_F}{\sqrt{2}\pi^2} \, m_t^2 \Lambda^2 \, ,
\ee
where $\Lambda$ is the upper limit of
the momentum in the loop-integration.
This correction diverges quadratically as $\Lambda$ is sent to infinity.
The renormalizability of the theory
 allows with a single subtraction to relate,  via a finite
relation, the Higgs mass parameter calculated at different scales:
\ba
m_H^2 (Q) &=& m_H^2(Q_0) + \nonumber \\
&& \frac{6G_F m_t^2 }{\sqrt{2}\pi^2} \,  (Q_0^2-Q^2) \; .
\ea
We say that the quadratic divergence is reabsorbed into the bare Higgs
mass parameter defined at the scale $Q_0$, $m_H(Q_0)$. This relation implies
that the combination
\be
m_H^2(Q_0) + \frac{6G_F m_t^2 }{\sqrt{2}\pi^2} Q_0^2
\ee
is a constant, independent of $Q_0$ for all values of $Q_0$ at which
the theory is represented by the SM. 

If we take $Q_0$ to be of the order of the EWSB scale, $v=247$ GeV,
and we use the range of \mh\  from the EW data, we
obtain for this constant a number of the order of few $\times 100$
GeV. If we allow $Q_0$ to become as large as the Planck mass $M_{Pl}
\sim 10^{19}$ GeV, the region where the SM gets unavoidably modified
by quantum gravity, $m_H^2(M_{Pl})$ must be fine tuned to the level of
$(v/M_{pl})^2 \sim 10^{-33}$ in order for the cancellation between
$M_{pl})^2$ and $m_H^2(M_{Pl})$ to result in a number of order $v^2$.
This fine tuning, while formally legitimate, is considered
theoretically to be extremely unnatural, and suggests to theorists
that eq.~\ref{eq:deltamh} should receive additional contributions
cancelling the quadratic term at
energy scales of $O$(few $\times v \sim $TeV), thus removing the
need for fine tuning. When theorists say that the SM is {\em
  incomplete}, they usually refer to this issue, called the
``hierarchy problem''. Most of the theoretical work of the past 30 years
has been devoted, directly or indirectly, to identifying solutions to
this problem. Supersymmetry, technicolour, large extra-dimensions, are
all different ways of addressing this issue. Their common approach is
to tie the Higgs boson to some new symmetry, which protects its mass
against the appearance of quadratic divergencies (see
\cite{Rattazzi:2005di} for a more complete discussion and for references). 

In supersymmetry this is achieved by introducing a fermionic
partner. Since fermion masses only receive logarithmic corrections,
the Higgs mass correction must be logarithmic as well. The way this
happens in practice is through the addition of the stop quark
$\tilde{t}$ (the
supersymmetric partner of the top) contribution to the
radiative corrections to $\mh^2$. The quadratic component of this contribution
has the same size of the top one, but opposite sign due to Bose
statistics, leading to a cancellation which leaves only a finite
term, proportional to the logarithm of the ratio of stop and top
masses.  

In the so-called {\em little-Higgs} theories, which are a modern
incarnation of technicolour, one introduces a global symmetry under shifts of
the Higgs field, $H\to H+a$. In this way, the fundamental Lagrangian
can only contain terms proportional to derivatives of the Higgs field,
and no mass can be present. When this symmetry is broken, only small
corrections to the Higgs mass can arise, and the radiative correction
are protected against the appearance of logarithmic contributions. In
these theories new particles are required to enforce this cancellation
at the diagrammatic level. In the case of the simplest little-Higgs
theories, these are new, heavier partners of the top quark, and new
gauge bosons $W'$ and $Z'$, all with masses in the 1--few TeV range.

In theories with extra dimensions, the Higgs is a component of gauge
fields along the extra dimensions, something that behaves as a scalar
in 4 dimensions. The gauge symmetry that protects the mass of gauge
bosons will then take care of eliminating the quadratic divergence,
using once again the contributions  to the Higgs mass loop corrections
of the  new particles appearing as Kaluza-Klein modes.

In all of these examples, care must be exercised to ensure that the
impact of the new particles on the EW observables be compatible with
the current precision measurements. This, together with the request
that the reduction of the fine-tuning is not spoiled by the
introduction of new very large mass scales, leads to the prediction of
a rich phenomenology of new phenomena at scales potentially within the
reach of the LHC. 

\section{The experimental hints for new physics}
While the above ideas are considered as sufficiently compelling by
most theorists to justify great optimism in the appearance of new
phenomena at the LHC, it is encouraging that also more pragmatic,
data-driven considerations, point in the same direction.
There are in fact at least three compelling
experimental observations that clearly demand new physics BSM: neutrino mixing,
dark matter, and the baryon asymmetry of the Universe, namely the
amount of baryons emerging from the early Universe. None of these
observations can be accommodated within the SM, regardless of how much
we allow ourselves to fiddle with possible uncertainties in the
theoretical predictions. Independently of our personal level of
pragmatism and indifference towards theoretical speculations such as
those presented in the previous section, as scientists we therefore have to
accept the existence of physics BSM. 

In addition to those three
clear cases, an increasing number of less significant,
but nevertheless tantalizing, indications of possible
discrepancies with the SM are emerging in various
low-energy observables. The crucial issues for the future of our field
are therefore the following. Is there a common thread among all
 deviations from the SM, pointing towards a new paradigm in
particle physics? If so, how soon, and with which experimental tools,
will it be possible to learn more about it?  
I personally feel that it is justified to answer positively to the
first question, and to expect that the field is ready, with the
forthcoming generation of experiments, to start unveiling and quantitatively
exploring these new phenomena.

I will now elaborate a bit more on this by using the example of
neutrino physics, and reviewing some of the weaker but nevertheless
interesting anomalies alluded to above.

\subsection{Neutrinos}
Neutrino masses themselves do not provide a new theory, and can be
incorporated within a trivial extension of the SM. What is exciting is
that once we look beyond this trivial realization in terms of sterile
right handed neutrinos, we find an amazingly fertile terrain for
interesting speculations: the connection with GUT-scale physics, via
the see-saw mechanism, is as strong as, if not stronger than, the
unification of the gauge couplings. The coincidence between these
two totally independent hints at grand unification certainly adds to
their individual strength! The failure of the SM to accommodate the
baryon asymmetry of the Universe makes leptogenesis (the lepton-driven B
asymmetry of the Universe) a very exciting possibility. The connection
with GUT, and the fact that SUSY quantitatively accommodates gauge
coupling unification much better than any non-SUSY GUT, strengthens
the case for SUSY itself. Willing to explore the broad consequences of
neutrino masses and to anticipate possible experimental needs to
analyze them, it is mandatory to explore the joint implications of
SUSY and neutrino masses and mixings. As briefly summarized here,
these are manifold and far reaching~\cite{Masiero}.

The form of the most general terms leading to neutrinos masses is
given by:
\ba
L_m &\propto& \, y_\ell \, H_{\ell} \, L_i \, L^c_i\nonumber \\
& +&  y^{ij}_\nu \,
H_{\nu} \, L_i \, N_j \; +\; M^{ij}_N \, N_i \, N_j 
\ea
If $M_N=0$, and the Higgs field coupled to $N$ is the conjugate of the
SM Higgs field, then we have a trivial extension of the SM, and the
smallness of the neutrino masses is driven by the ({\em unnatural?})
smallness of the Yukawa couplings $y_\nu$. In this scheme neutrino
masses and mixings are given parameters, without any dynamical
content. Neutrino mixings will lead to FCNC processes in the charged
lepton sector. Their rates will be proportional to the probability
that a neutrino can oscillate and change flavour during the time
allowed to the virtual transition $\ell \to \nu_\ell W \to \ell$,
leading to invisibly small  effects. The presence of the additional
Majorana mass term enables the see-saw mechanism: a large value of
$M_N$ leads to the following solution for the light eigenvalues of the
$(\nu,N)$ mass matrix:
\be
M_\nu = -y_\nu \,  M_N^{-1} \, y_\nu^T \; \langle H_\nu \rangle^2
\ee
Assuming that the expectation value of the Higgs field coupling to
neutrinos is of the same order of magnitude as that of the SM Higgs,
one can generate masses in the range given by data provided $M_N$ is
of order $10^{15}$ GeV. In the presence of Supersymmetry, $H_{\nu}$ is
not an arbitrary new Higgs field, but is the Higgs giving mass to the
up-type quarks. Taking seriously the connection with the GUT scale
suggested by the see-saw mechanism, enforcing the GUT symmetry on the
high-energy lagrangian leads to an even more direct relation between
up-type quark and neutrino masses. For example, in the simplest case
of a $SO(10)$ theory we would have:
\ba
L_m & \propto& \, y_{i,j}^{d,\ell} \, {\mathbf{16}}_i {\mathbf {16}}_j
H^{10}_{d}  \nonumber \\
&+&  y_{i,j}^{u,\nu} \, {\mathbf {16}}_i {\mathbf {16}}_j
H^{10}_{u} \; + \; \, y_{i,j}^{R} \, {\mathbf {16}}_i {\mathbf {16}}_j
H_{R}^{126} 
\ea
where the Higgs fields coupling to up and down-type quarks $H_u$ and
$H_d$ lie in different 10-dimensional representations of $SO(10)$, and 
\be
\mathbf{16}=(u_L,d_L,u^c,e^c)_{10} + (d^c,L)_{\bar{5}} + N^c
\ee 
is the $SU(5)$ decomposition of the representation containing all SM
fermion fields plus the left-handed anti-neutrino $N^c$.  The first
consequence of these relations is that at least one entry in the
neutrino mass matrix is of the order of the top Yukawa coupling, and 
\be
m(N)\sim m_{top}^2/m_{\nu}
\ee
for the third-generation neutrinos. 
\begin{figure}[htb]
\includegraphics[width=0.45\textwidth]{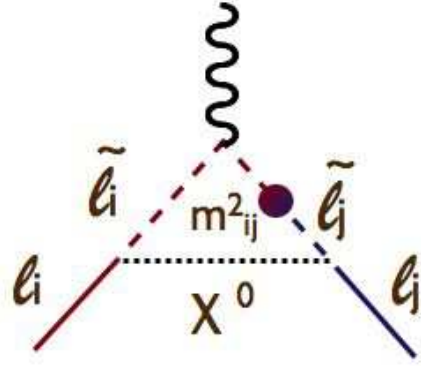}
\caption{$\ell_i \to \ell_j\gamma$ decay induced by the mixing of
  charged sleptons.}
\label{fig:triangle}
\end{figure}
Assuming that $m(N)$ should not
exceed the GUT scale, there is a lower limit on the mass of the light
3$^{rd}$-generation neutrino. The second consequence is that quark mixings
lead to charged slepton mixing via renormalization group evolution
from $M_{GUT}$ to $m_N$. In the case of supersymmetry breaking induced
by common soft scalar masses $m_0$ at the GUT scale, one for example obtains: 
\be
(m^2_{\tilde L})_{ij} \; \sim \; -\frac{3m_0^2+A_0^2}{8\pi^2} \; y_t^2
\; O_{ij} \; \log\frac{M_{GUT}}{M_{N_R}} ; ,
\ee
where
\be
y_t^2 \, O_{ij} = \sum_{k} \; y^{\nu}_{ik} \,y^{\nu *}_{kj}
\ee
\begin{figure}[htb]
\includegraphics[width=0.45\textwidth]{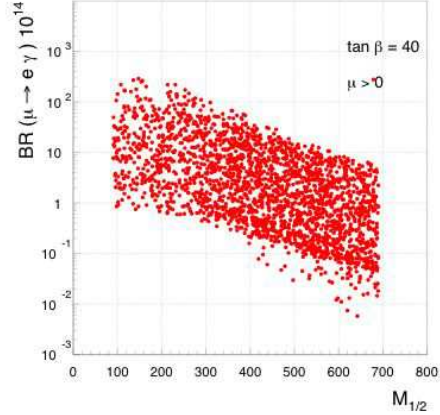}
\caption{$\mu \to e\gamma$ decay rates in the CKM case.}
\label{fig:mueg}
\end{figure}
\begin{figure}[htb]
\includegraphics[width=0.45\textwidth]{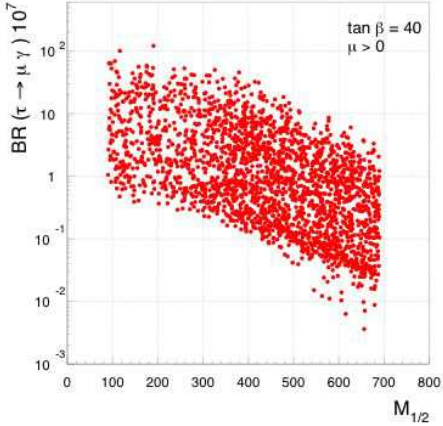}
\caption{$\tau \to \mu\gamma$ decay rates in the PMNS case.}
\label{fig:taumug}
\end{figure}

\begin{figure}[thb]
\includegraphics[width=0.45\textwidth]{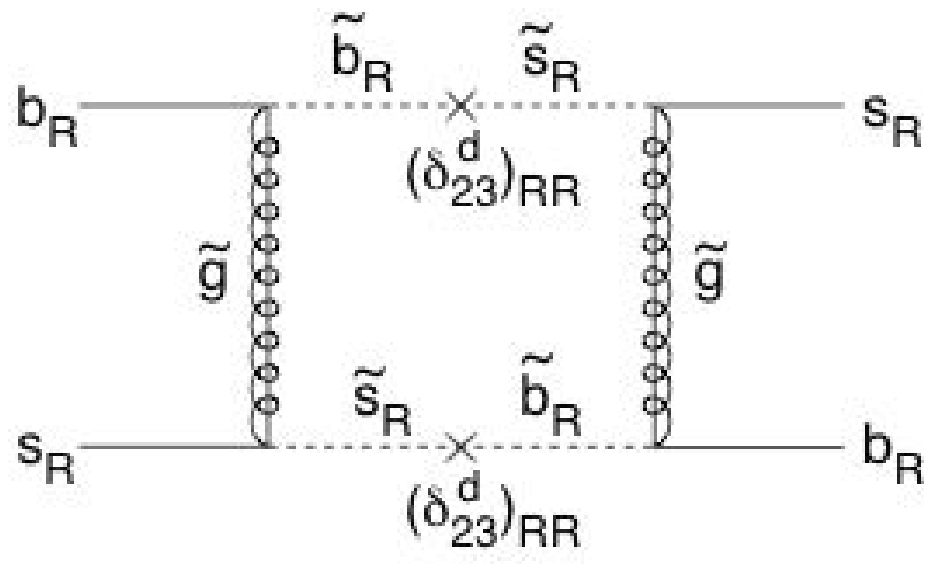}
\includegraphics[width=0.45\textwidth]{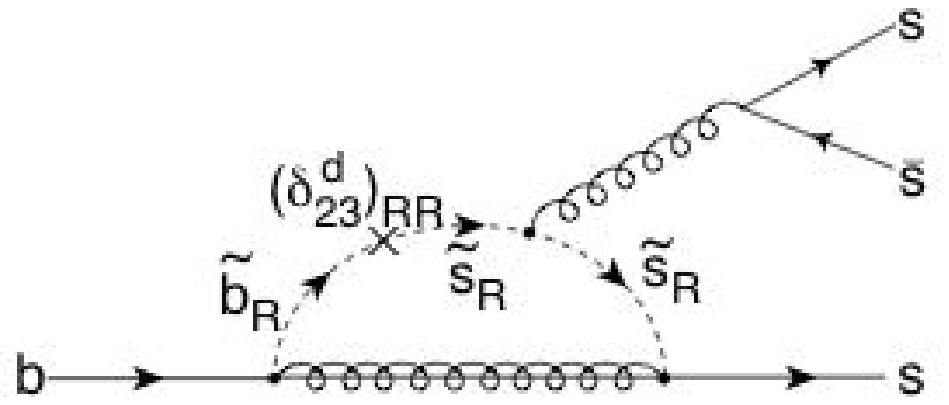}
\caption{Possible contributions to $B_s$ mixing and CP violation in $B\to
 \psi\phi$ and $B\to \phi K_s$, 
induced by mixing among
  ``right-handed'' $b$ and $s$ squarks.}
\label{fig:bsmix}
\end{figure}

The diagrams in fig.~\ref{fig:triangle} will then induce transitions
such as $\mu\to e\gamma$ or $\tau \to \mu \gamma$. The quantitative
prediction for the branching ratios depends on the specific values of
the entries of the mass matrix $O_{ij}$, something that the available
data cannot allow to uniquely fix. One could envisage two cases as
representing the possible range: \be O_{ij} \propto \sum_{k} \;
V^{CKM}_{ik} \,V^{CKM *}_{kj} \; , \ee (the CKM case) where $V^{CKM}$
is the CKM matrix, and \be O_{ij} \propto \sum_{k} \; U_{ik}
\,U^{*}_{kj} \; , \ee (the PMNS case) where $U$ is the neutrino mixing
matrix. In the CKM case the transitions $\mu\to e\gamma$ or $\tau \to
\mu \gamma$ are proportional to $\vert V_{ts}V^*_{td} \vert^2$ and
$\vert V_{tb}V^*_{ts} \vert^2$, respectively. One therefore expects the
$\tau\to\mu$ decay to have a BR a couple of orders of magnitude larger
than $\mu\to e$.  In spite of the smallness of the CKM matrix
elements, rates for $\mu\to e\gamma$ can still be large enough to be
within the reach of the forthcoming  experiments for a
large fraction of the model parameters space, as is shown by the
scatter plot in fig.~\ref{fig:mueg}.  In the PMNS case, the rates can
be significantly larger, given the larger size of the neutrino mixing
matrix elements. In the particular case of $\tau\to\mu$ transitions,
the rate is proportional to $U_{\mu 3} U_{\tau 3}$, which is known and
large. A large fraction of parameter space would already be excluded
by the current limits~\cite{Aubert:2005wa}, in the range of $BR(\tau
\to \mu \gamma) < 10^{-7}$, as shown in fig.~\ref{fig:taumug}.  The
$\mu \to e$ transition depends on $U_{e 3}$, which is yet unknown and
could be very small.

There is another important by-product of SUSY-GUT frameworks for
neutrino masses, which leads to possible manifestations in the quark
sector: since the neutrinos sit in the same $SU(5)$ multiplet as
down-type antiquarks (or right-handed quarks), the large mixing
between $\mu$ and $\tau$ neutrinos leads to a large mixing between
right-handed quarks.  This has no impact on phenomenology, since
right-handed quarks do not couple to weak interactions. However it
feeds via Supersymmetry into a large mixing between the scalar
partners of R-handed squarks, and to interactions like the ones shown
in fig.~\ref{fig:bsmix}. The first contribute to $B_s$
mixing and possibly CP violation in $B\to\phi \psi$ decays (which is
approximately 0 in the SM). The second lead to an extraction of $\sin
2\beta$ from $B\to \phi K_S$ decays which is different from what
obtained in $B\to \psi K_S$. 

\subsection{Clues from flavour physics}
We tend to associate today the origin of the SM with the gauge
principle and the consolidation of Yang-Mills interactions as unitary
and renormalizable quantum field theories. We often forget
that flavour phenomena have contributed as much as the gauge
principle, if not more, in shaping the overall structure of the SM. It
is the existence of flavours (both in the lepton and quark sector)
which gives the SM its family and generation structure. The idea of assembling
quarks in EW doublets was guided by the suppression of FCNCs, which
led to the GIM mechanism and to the prediction of the charm
quark. Kaon decays led to the observation of CP violation, and to the
CKM model. $B_d$ mixing, similarly to the role played by
$K^0-\bar{K}^0$ mixing in getting the mass range for charm, was the
first experimental phenomenon that correctly anticipated the large
value of the top quark mass. And, last but not least, the observation of
neutrino masses provides today the first concrete and uncontroversial
evidence that the SM is incomplete: most modestly, we need to introduce
degrees of freedom for sterile right-handed neutrinos; more ambitiously,
neutrino masses are a window on physics at the grand unification scale!

\begin{figure}[htb]
\includegraphics[width=0.45\textwidth]{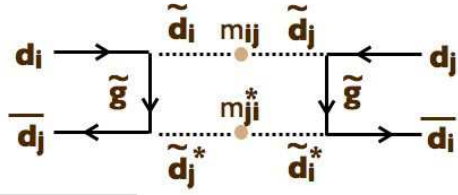}
\caption{Contribution to $Q-\bar{Q}$ mixing ($Q=K,B$) from gluino and
  squark exchange in Supersymmetry.}
\label{fig:box}
\end{figure}

The smallness of FCNC and of CP violation are not an outcome of the
SM: they have been built into its structure from the outset.  In the
quark sector, they are guaranteed by the unitarity of the CKM mixing
matrix and by the small mixing between heavy and light generations.
The transition $d_i \bar{d}_j \to X$, where $d_i$ are down-type
quarks, $i\ne j$ and
$X=\ell^+\ell^-$ or $d_j \bar{d}_i$, is proportional to 
\ba
\Delta_{ij} &\sim & \sum_{k=u,c,t} \; V_{ki} \, V^*_{kj} \;
f(m_k/m_W) \nonumber
\\
&\sim& \sum_{k=c,t} \; V_{ki} \, V^*_{kj} \; m^2_k/m^2_W \; ,
\ea
where $V_{ij}$ are the elements of the CKM mixing matrix. 
As a result of the unitarity of $V_{ij}$, the leading contributions to
this expression are given by
\be
\; V_{ci} \, V^*_{cj} \, \frac{m_c^2}{m_W^2} \,+\, V_{ti} \,
V^*_{tj}  \; .
\ee   
The first term is strongly suppressed by the charm mass (GIM), the
second by the smallness of the mixing of the third generation with the
first two. In the lepton sector, it is the smallness of the neutrino
masses that suppresses possible evidence for mixings and CP violation
for charged leptons.
There is absolutely no guarantee that the above properties survive in
extensions of the SM~\cite{Buras:2005xt}. 
A typical example of what may happen is given in
fig.~\ref{fig:box}: if the squark mixing matrix is not aligned with
CKM, the squark mass eigenstates can mix and lead to potentially large
contributions to $K\bar{K}$ or $B\bar{B}$ mixing. In a model where
squark flavours are maximally mixed, squark and gluino masses should
be larger than several TeV in order to sufficiently suppress these
contributions and not clash with the data on
mixing or CP violation in the $K$ sector! 
As long as no evidence is brought forward for the existence of
supersymmetry, this is not an issue. The day that supersymmetry (or
other forms of new physics) should be discovered at mass scales below
or around the TeV, say at the LHC, 
understanding how this problem is bypassed will become one of the
the most exciting issues in our field!

\subsection{Hints of more to come}
Before that day arrives, new very accurate data from flavour physics
start providing interesting and tantalizing clues for the existence of
small deviations. While still not significant from the
statistical/systematic viewpoint, these deviations are enoguh to
keep our expectations high.    

In addition to the well known discrepancy~\cite{Bennett:2006fi}
 in the anomalous magnetic moment of the muon, $a_\mu=(g-2)_\mu$:
\be
a_\mu^{SM} - a_\mu^{exp}=(2\pm1)\times 10^{-9} \; ,
\ee
two recent long-awaited measurements have appeared, both indicating a
 deviation from the SM at the level of $\sim 2\sigma$~\cite{Bona:2006ah}. 
The first is the
 determination of the $B_s$ oscillation frequency by 
CDF~\cite{Abulencia:2006mq}:
\ba
&&
\Delta M_{B_s}(\mathrm{ps}^{-1}) = \nonumber \\
&& (17.31 {+0.33 \atop -0.18}_{stat} \pm 0.07_{syst})
 \, ,
\ea
which is slightly smaller than the SM expectation:
\be
\frac{\Delta M_{B_s}^{exp}}{\Delta M_{B_s}^{SM}} = 0.80 \pm 0.12 \, .
\ee

The second is the detection by Belle of the $B\to \tau\nu$
decay~\cite{Ikado:2006fc}\footnote{A measurement by BaBar recently
  appeared as well~\cite{Aubert:2006fk}, 
$B_\tau=(0.88^{+0.68}_{-0.67}({stat.}) \pm 0.11 ({syst.})) \times 10^{-4}$.}, 
with a branching ratio
$B_\tau=B(B^-\to \tau^-\nu)$ measured as:
\ba
 && 10^4 \times \btau= \nonumber \\
 &\quad &(1.06 {+0.34 \atop -0.28}_{stat} {+0.18 \atop
  -0.16}_{syst})
\ea 
again slightly lower compared to the SM prediction (normalized here to
the $B_s$ mixing rate, to reduce the theoretical
systematics~\cite{Isidori:2006pk}):
\ba
&& \frac{\btau^{exp}/\Delta M^{exp}_{B_s}}{\btau^{SM}/\Delta
  M^{SM}_{B_s}} = 0.67  \\
&& {+0.27 \atop -0.22}_{exp} \pm 0.06_{\hat B_{B_d}}  \pm
0.07_{\vert V_{ub}/V_{td} \vert } \nonumber \, .
\ea
These deviations have been found by several
authors (see e.g.~\cite{Isidori:2006pk,Blanke:2006ig,Lunghi:2006uf}) 
to be consistent with a SUSY
extension of the SM, with a large value of the Higgs mixing angle
$\tan\beta$ and with charged Higgs bosons with mass in the range of few hundred
GeV. Such scenarios will be well explored by the LHC and the ILC by
directly producing and observing the new states required.

Potentially measurable effects could also emerge in a violation of
universality~\cite{Masiero:2005wr} in
\be
R_K=\Gamma(K\to e \nu)/\Gamma(K\to \mu\nu) \; .
\ee
Expressing any deviation from the SM as
$R_{K,\pi}=R_{K,\pi}^{SM}(1+\Delta r^{e-\mu}_{K,\pi\; NP})$
one has, for $\Delta_R^{31}\sim 5\times 10^{-4}$ (the parameter
proportional to the mixing between the 1$^{st}$ and 3$^{nd}$
generation sleptons), $\tan\beta=40$
and $m_{H^\pm}=500$ GeV~\cite{Masiero:2005wr}:
\ba
&&\Delta r^{e-\mu}_{K,\pi\; NP} \sim \nonumber \\
&& \left(\frac{m_K^4}{m_{H^\pm}^4}\right)
\left(\frac{m_\tau^2}{m_e^2}\right) \vert \Delta_R^{31}\vert^2 \,
\tan\beta^6 \nonumber \\
&&\sim 10^{-2} \; .
\ea
At this time, NA48/2 sets the bound~\cite{Ceccucci:2006mk}:
\be -0.063 < \Delta r^{e-\mu}_{K,\pi\; NP} < 0.017
\ee
with a theoretical uncertainty at the per mille level. Future
experiments should be planned to match this accuracy.

\section{Conclusions}
Progress in the field will be essentially driven by new and better
experimental data.  Theorists have pretty much exhausted their arsenal
of weapons to make progress based on first principles
only. Nevertheless, they have created scenarios for BSM physics that,
in addition to addressing the most outstanding theoretical puzzles and
the established deviations from the SM (DM, BAU, $\nu$ mixing), predict
galore of new phenomena at energy scales and precision levels just behind the
corner.
If the only open questions required experiments at the GUT scale, HEP might be
stuck for a long while.  Fortunately, both theoretical and
experimental issues point instead at the TeV scale.  There is
therefore a solid and justified expectation that progress will start
emerging from the forthcoming generation of experiments, for which the
detectors are being completed. This progress will bring a major
revolution in HEP.  It will not be a minor adjustment or incremental
progress, like may have been the discovery of a 3$^{rd}$ generation. We
expect to uncover qualitatively new phenomena (e.g. SUSY), which will
lead to a quantum leap in our understanding of the Universe, and will
open new prospects for experimental research.

We have two main sets of laboratory tools available:  the high-energy
frontier (LHC, ILC, etc), which we anticipate will mostly address the
problem of EWSB, and therefore ``gauge-sector'' issues; and  the
low-energy, high-intensity frontier, where experiments will probe
mostly the ``flavour sector'' (flavour-changing transitions and CP
violation phenomena with quarks, neutrinos and charged leptons).  We
still don't know which of these two directions will provide more
fruitful. Most likely, it will be the complementarity between these
two experimental approaches that will give us most insights.
Whether or not new physics is seen at the LHC, maintaining diversity in the 
experimental programme is therefore our best investment for HEP.
 If new physics (especially SUSY) is discovered at the LHC, a global flavour 
physics programme (LFV and CP/FCNC in the quark sector) is an 
essential component of the HEP research, required to explore the nature of 
the new BSM framework (e.g. to identify the SUSY breaking scenario). 
An ambitious and far-sighted $\nu$ programme is likewise a mandatory element of
the HEP future, as this programme has concrete
 goals and benchmarks, and a direct
impact on our ability to uncover new information about nature: GUT,
CPV, BAU.  Furthermore, as the indirect evidence for GUT grows, proton decay
searches should continue.

The political feasibility of such a broad research programme was
addressed at this meeting during the round table by the Directors of
laboratories, funding agencies and the leaders of ongoing global
projects.  The experimental feasibility, in terms of suitable
detectors, is what this conference is all about!

Exciting times are ahead of us, and the least we can do is to get
ourselves ready on all fronts to face the challenge!

\end{document}